\documentstyle[12pt]{article}
\title
{SHELL EFFECTS IN NUCLEI NEAR NEUTRON DRIP}
\author{M.M. SHARMA$^1$, G.A. LALAZISSIS$^2$, W. HILLEBRANDT$^1$,
P.RING$^2$ \\
$^1$Max Planck Institut f\"ur Astrophysik,\\
Karl-Schwarzschildstrasse 1, D-85740 Garching, Germany\\
$^2$Physik Department, Technische Universit\"at
M\"unchen\\
 D-85747 Garching, Germany}
\begin {document}
\maketitle
\begin{abstract}
Shell effects in nuclei close to the neutron-drip lines have been
investigated. It has been demonstrated in the relativistic mean-field
theory that nuclei very far from stability manifest the shell
effects strongly. This behaviour is in accord with the  predictions
of nuclear masses in the finite-range droplet model including
shell corrections. As a consequence we find a large neutron halo in
nuclei near the neutron-drip line. The shell effects predicted
in the existing Skyrme mean-field theory in comparison
are significantly weaker than those of the other approaches.
\end{abstract}
\newpage
Very neutron-rich nuclei and, in particular, those near the neutron-drip
lines and near closed shells, play an important role in nuclear astrophysics.
Their properties such as binding energies, neutron separation energies,
deformation parameters, etc., strongly affect the way neutron-rich
stable isotopes are formed in nature by the so-called r-process.
Conversely, a precise knowledge of such properties will help to
determine the astrophysical conditions of their formation [1].
A second example is the equation of state of neutron stars, where
in the crust neutron-rich nuclei persist in coexistence with a gas
of free neutrons and in the core the isospin dependence of the
nucleon-nucleon interaction determines the stiffness of the equation
of state and, consequently, their maximum mass, moment of
inertia, etc.[2,3].

The shell effects in nuclei manifest themselves in the
form of magic numbers. These depend principally on the spin-orbit
coupling and seem to have been understood along the valley of
$\beta$-stability. Theoretically, description of nuclei is provided
by the density-dependent Skyrme mean field [4], which has
been employed for two decades, and by the relativistic
mean-field (RMF) theory [5,6]
a framework which has been successful for
finite nuclei. Whereas in the former ground-state properties
of nuclei close to the magic numbers can be  described successfully,
the properties of nuclei away from stability still remain
a problem. In the RMF theory, on the other hand, the binding energies,
charge radii and neutron-skin thickness of nuclei with closed-shells
as well as of those far away from stability can be reproduced
well [7]. Approaches based upon semi-empirical
mass formulae [8]
with a large number of parameters also strive to provide
descriptions of nuclei close to and far off the stability line.
Although many of these mass formulae give similar fits along the line
of stability, their predictions vary considerably for short-lived
nuclei and especially near the drip line.

In this letter, we examine in the RMF theory the behaviour of the shell
effects in nuclei near neutron drip. We show that for the first time
a microscopic model describes the properties of nuclei very far off
stability, such that the results are in striking conformity with
predictions of the empirical finite-range droplet model [9].
We also show that they differ considerably from
those provided by the Skyrme  mean-field approach.

We have considered the chain of heavy Zr isotopes to demonstrate the
shell-effects. Nuclei from A=110 to A=136 with even neutron number
have been included. This range is well
outside the valley of $\beta$-stability and encompasses also the
neutron-drip line. The shell-closure occurs at A=122 (N=82).
This nucleus lies close to the neutron-drip line. In the nuclei
heavier than this, shell effects due to the next higher shell might
come into play.

Calculations have been performed with the relativistic Hartee approach
taking deformed configuration into account. The force NL-SH
(ref. [7]), which has been shown to describe the properties of
nuclei at and away from the
stability line very well, has been used. The correct asymmetry energy
of NL-SH allows its judicious application to nuclei with large neutron
excesses. For open-shell nuclei, pairing has been added in the BCS formalism
with constant pairing gaps.

Figure 1 shows the ground-state binding energies of Zr isotopes
obtained with NL-SH. With each successive addition of a pair of neutrons,
the nuclear binding energy increases monotonously from A=110
to A=122 where the shell closure occurs. Further
addition of neutrons to the magic core puts the extra
neutrons in the higher shell. Evidently, these neutrons
provide barely any extra binding to the nucleus and the binding
energy in this region stagnates. This depicts an onset of the
neutron-drip where the outer neutrons become loosely bound.
The kink in the binding energy curve at A=122  signifies
the strong shell-effects that are encountered when further
neutrons are added to this nucleus. Calculations assuming
spherical configurations for all Zr isotopes do yield a
curve very close and similar in shape to the deformed case and the
shell effects are thus retained also in the spherical case.

In Fig. 1 we give also a comparison with the finite-range droplet
model [9] (FRDM). The agreement of our results with
those of the FRDM is striking. From A=110 to A=128 the deviations
are atmost 1-2 MeV. The shell effects in both the RMF theory and
the FRDM are amazingly similar. Even beyond A=130 there are only
minor differences between the two and the
binding energies differ by about 2-3 MeV. The general trend in the
stagnation in the binding of nuclei close to the drip line is
exhibited excellently by both the RMF theory and the FRDM.

For comparison we include in Fig. 1 the binding energies calculated
within the spherical Skyrme Hartree-Fock and BCS approximation using
the interaction SkM* [10]. The SkM* energies deviate strongly
from FRDM and NL-SH
as the neutron number increases. The lack of shell effects at N=82 in
the Skyrme approach is evident from the smooth variation in energy
about A=122. This seems to be a typical feature of Skyrme forces,
which has also been observed in ref. [11] in several chains of
isotopes crossing shell-closure.

Finally, we give in Fig. 1 also the results for the  force NL1. [6]
The binding energies for NL1 differ from NL-SH and FRDM by upto 15 MeV.
This is due to very large asymmetry energy of 44 MeV of the force NL1,
which also leads to neutron skin-thickness [12] larger than the
empirical values. The kink at A=122 present also in NL1, however, points
again to shell effects similar to NL-SH. Such kinks, thus, seem to be
a characteristic of the RMF theory in contrast to the Skyrme ansatz.

Deformation properties of nuclei have been obtained from the RMF
theory. The quadrupole deformation $\beta_2$ from
NL-SH is shown in Fig. 2. The values obtained from the FRDM
are also  shown for comparison. Both the RMF theory and the FRDM
agree remarkably well on the shape transition from prolate to oblate
at A=114. The $\beta_2$ values from the RMF theory are in general
close to those from the FRDM. The fact that nuclei above
A=126 again assume prolate shape in the FRDM is well reproduced
by NL-SH. Only above A=116 the shape transition from oblate to
spherical in approaching the closed-shell is more gradual in the RMF
theory than in the FRDM. The agreement in the deformation properties of
nuclei near the drip line predicted by both
the RMF theory and the FRDM is astonishing.

In order to understand why the binding energies beyond the closed neutron
shell stagnate, we show in Fig. 3 single-particle energies of levels below
and above the Fermi energy for both (a) NL-SH and (b) SkM*.
The noteworthy aspect in this figure is the change in the Fermi surface
as a function of neutron number, as shown by the dotted curves.
Both the approaches show a sudden change in the
Fermi surface as the next shell is being filled. The change in the
Fermi surface after A=124 for NL-SH is very gradual and as the neutron
number increases, it is connected very gradually to the continuum. In this
process the nuclei still take up more neutrons without contributing to
the total binding energy. This also is the reason that the drip-line
is approached considerably earlier than with SkM* and that
already above A=124 neutrons get unbound. This picture contrasts
with the one of SkM*, where the Fermi energy reaches the continuum limit
only for very heavy isotopes, leading to an increase in the binding energy
with addition of further neutrons to the next shell.
The drip line is encountered suddenly at about A=134 and
therefore the coupling to the continuum arises rather suddenly. The drip
nucleus being close to A=134 is consistent with the predictions
of ref [11], where with the Skyrme force SkP shell-effects
were found to be very weak for extremely neutron-rich nuclei.

Shell effects in the RMF theory and the Skyrme approaches have
recently been a matter of discussion for stable nuclei also.
Experimentally, isotope shifts in nuclei near shell-closure have
always been found to show a kink [13]. Such kinks can be
construed as being a manifestation of shell effects. Microscopic
calculations have until recenly not been successful in describing such
kinks. The Skyrme mean-field approach [14] including various
other correlations has failed to reproduce such empirical effects.
Only very recently, the RMF theory has succeeded [15] in describing
these kinks in charge radii differences.
The kinks in the binding energies in Fig. 1 in the RMF theory and the
FRDM reflect shell effects which may be of similar nature as in the
charge radii. However, here the issue concerns the shell effects
at the drip line which is truly far away from the stability.

Shell effects at drip lines have also been looked into in ref. [16]
using the Skyrme HFB and the RMF theory. Contrary to clear
differences found in our work between the Skyrme and the
RMF approaches, ref. [16] points to a quenching of
shell effects in both the theories, a conclusion which is clearly
in contradiction with ours.  Variations in pairing and single-particle
properties with mass number along an isobaric chain about A=100
have been shown [16] to be similar both for the Skyrme and
the RMF theory, a situation very different from our
figure 3, although the weakening of shell-effects near the drip-line
with the Skyrme interaction SkP (ref. [16]) is consistent with
that of ref [11]. Using the same interaction, this result
obviously is inconsistent with the shell effects exhibited by our
results in the RMF theory. The spin-orbit splitting of some levels
in both the approaches are considerably different (Fig. 3).
We suspect that the differences between our conclusions and those of
ref. [16] in part stem from the selection of isobaric chains
in [16], which does not include any shell-closure.
and therfore, the shell effects do not manifest themselves clearly.
Indeed, in the middle of a shell both the approaches
provide similar results as shown in ref.[16].

We have compared some of the qunatities that may lead to the differences
in the shell effects in the two approaches.
The sum of single-particle energies  in the two approaches, both for
the protons and neutrons, do not show any compelling
difference as do also the interaction
energies as a function of mass (the details will be provided elsewhere).
The differences between the two, however, become evident when we examine the
single-particle structure and the corresponding spin-orbit
splittings. The orbits 1$g_{7/2}$ and 1$h_{11/2}$ seem to play a dominant
role in defining the shell effects in the two approaches, whence
the differences appear. First, the spin-orbit splittings
1$h_{9/2}$-1$h_{11/2}$ and 1$g_{7/2}$-1$g_{9/2}$  are  systematically
larger in the Skyrme
approach than in the RMF theory by 30\% and 20\%,
respectively, as can be seen partly from Fig. 3. The splitting
2$d_{3/2}$-2$d_{5/2}$ for SkM* is about 2 times larger than the one
for NL-SH. Thus, the larger spin-orbit splitting in
the Skyrme theory would tend to smoothen the shell effects.
Secondly, the slightly different sequence of single-particle
levels arising from differences in the spin-orbit splitting
lead to different filling of the orbitals. As the mass number
increases from A=110 to A=122, the preferred subshell to be filled
in the RMF theory is 1$g_{7/2}$, which lies deeper than in SkM*,
whereas in the Skyrme theory 1$h_{11/2}$ is being filled at the
expense of 1$g_{7/2}$. The different occupations of levels give
rise to different Fermi energies which in turn modify the shell effects.

The differences in the spin-orbit properties of the two approaches
can be traced back partly to the density-dependence of the spin-orbit
term. In the Skyrme ansatz, it takes the form:

\begin{equation}
{3\over 4}W_0 \Bigl({1\over r}{d\over dr}\rho\Bigr)\vec\sigma.\vec L,
\end{equation}

where $W_0$ is the spin-orbit strength. A non-relativistic reduction
of the relativistic Hamiltonian, on the other hand, gives rise
to an approximate spin-orbit term:

\begin{equation}
{1\over m^{*2}} \Bigl({1\over r}{d\over dr}\rho\Bigr)\vec\sigma.\vec L
\end{equation}

It can be seen clearly that the coefficient of the spin-orbit term
in the RMF theory possesses a density-dependence which goes as
$1/m^{*2}$ in contrast to $W_0$ in the Skyrme theory, which is indepedent
of $\rho$. This density dependence in the RMF theory arises as a
natural consequence of the Lorentz
structure of the interaction, where the saturation mechanism
is different from that of the Skyrme forces. A detailed
study on these aspects is in progress.

The consequence of the above Fermi surface in the RMF theory is evidently
a large tail in the density of the neutrons, i.e., further
addition of neutrons to the A=122 core leads to an extension of
neutrons in the space. Thus, in the vicinity of the drip-line,
nuclei with a "giant halo" are being created.

In conclusion, we have shown that in the RMF theory the shell-effects at
the neutron-drip line are  significantly stronger than those in the
Skyrme mean-field theory. For the RMF theory, this view is supported
by the phenomenological finite-range droplet model, the overall
predictions of which are reproduced  very well by the RMF
theory. Striking differences have been found in the spin-orbit
properties of the RMF theory as compared to the Skyrme ansatz.
As a consequence we predict large neutron halos in nuclei near the
drip line. From our analysis we expect, furthermore, strong shell effects
also for other elements and magic neutron numbers in the r-process path,
in contrast to what was found in ref. [16]. This would be in
accord with the fact that strong r-process abundance peaks are observed
at A $\simeq$ 80, 130, and 195, which require shell effects for
N = 50, 82, and 126, respectively, and neutron separation energies
around 2 or 3 MeV [1].

One of the authors (G.A.L.) acknowledges support from the Deutsche
Akademische Austauschdienst (DAAD). M.M.S. would like to thank Chris
Pethick for useful discussions.

\newpage

{\bf figure 1}

The binding energy of Zr isotopes obtained from the
deformed relativistic Hartree calculations with the force [7] NL-SH
with constant pairing gaps obtained from $\Delta_{n(p)} = 12/\sqrt{N(Z)}$.
Results for NL1 [6] are also shown. The smooth curve shows
the binding energy from Skyrme interaction SkM* [10] using spherical
configuration. A comparison has also been made with the predictions
of the finite-range droplet model (FRDM) [9].

{\bf figure 2}

The quadrupole deformation $\beta_2$ obtained from
NL-SH. The values from predictions of FRDM are also shown. The
overall agreement of NL-SH with FRDM, including the shape transition
from prolate to oblate at A=112 is noteworthy.

{\bf figure 3}

The single-particle levels for (a) the RMF theory
and (b) the Skyrme ansatz. The Fermi energy in both the cases is
shown by the dotted curve.

\end{document}